# Two dimensional twisted chiral fermions on the lattice.

Rajamani Narayanan and Herbert Neuberger *

*School of Natural Sciences, Institute for Advanced Study,
Olden Lane, Princeton, NJ 08540*

## Abstract

It is shown that the lattice overlap correctly reproduces the chiral determinant on a two dimensional torus in the presence of nontrivial background Polyakov loop variables.

---

* Permanent Address: Department of Physics and Astronomy, Rutgers University, Piscataway, NJ 08855-0849

The ultimate aim of the recent efforts to construct chiral gauge theories on the lattice is to perform numerical simulations in four dimensions. For technical reasons, many of the preliminary tests are carried out in two dimensions [1]. In the context of string theory the two dimensional formulation may be of interest in itself. Here, we wish to report on a new two dimensional continuum test that the lattice overlap of ref. [1] passes successfully. After describing the test we shall explain why this test was a worthwhile check.

The problem studied here is the computation of the determinant of the chiral Dirac operator describing a left handed Weyl fermion on a torus in the background of a uniform $U(1)$ gauge field. The continuum answer is known [2] and we show that the lattice overlap reproduces it.

In the continuum we view the torus as the complex plane restricted by the identifications

$$x_1 + ix_2 \equiv z \sim z + 2\pi n_1 l_1 + 2\pi i n_2 l_2, \quad n_{1,2} \in \mathbf{Z}.$$

The chiral Dirac operator is

$$D = \partial_1 + i\partial_2 + i(A_1 + iA_2) \equiv \partial_z + iA,$$

where $A$ is the uniform background. $D$ acts on functions $\psi$ obeying anti-periodic boundary conditions,

$$\psi(x_1 + l_1, x_2) = \psi(x_1, x_2 + l_2) = -\psi(x_1, x_2).$$

By a similarity transformation, $D$ is equivalent to $\partial_z$ acting on functions $\tilde{\psi}$ obeying twisted boundary conditions,

$$\tilde{\psi}(x_1 + l_1, x_2) = -e^{iA_1 l_1}\tilde{\psi}(x_1, x_2), \quad \tilde{\psi}(x_1, x_2 + l_2) = -e^{iA_2 l_2}\tilde{\psi}(x_1, x_2).$$

The twisted boundary conditions do not distinguish between gauge potentials $A_\mu + \frac{2\pi k_\mu}{l_\mu}$ differing by $k_\mu \in \mathbf{Z}$, $\mu = 1, 2$. In terms of $D$, these potentials are distinct but related by periodic gauge transformations. If $\det D$ were gauge invariant it would be periodic in the $A_\mu$ with periods $\frac{2\pi}{l_\mu}$. As is well known, in the context of quantum field theory, $\det D$ cannot be made gauge invariant although $\det(D^\dagger D)$ can. Since $D$ is holomorphic in $A$, one may pick in some contexts a holomorphic definition for $\det D$, $\chi(A)$. However, any gauge invariant definition of $\det(D^\dagger D)$ cannot be written as $|\chi(A)|^2$.

Introduce the variables

$$\tau = i\frac{l_1}{l_2}; \quad \alpha = \frac{l_1 A_1}{2\pi} + \tau \frac{l_2 A_2}{2\pi} = \frac{l_1}{2\pi}A.$$

We require $\det D$ to obey the following:

1. Invariance under a 90 degrees rotation corresponding to $\alpha \to \alpha' = \alpha/\tau, \quad \tau \to \tau' = -\frac{1}{\tau}$.
2. Invariance under charge conjugation, $\alpha \to -\alpha$.
3. Parity violation should be restricted to the phase of $\det D$, i.e.

$$\det D(\alpha^*, \tau) = (\det D(\alpha, \tau))^*.$$

4. Gauge invariance of $|\det D|$, which means

$$|\det D(\alpha, \tau)| = |\det D(\alpha + 1, \tau)| = |\det D(\alpha + \tau, \tau)|.$$

One can think about $\log \det D$ as being given by the usual sum over external legs attached to a single chiral fermion loop. The freedom in defining $\log \det D$ reflects the $\alpha$ dependence of the divergent graphs and is therefore restricted to an $\alpha$ independent term and a quadratic one. The $l_1, l_2$ dependence of the quadratic term is trivial since the renormalization ambiguity is strictly local. Starting from any acceptable form of $\det D$, one finds that the $\alpha$ dependent freedom is fixed by the imposition of $1 - 4$, and the result can be written as

$$\det D(\alpha, \tau) = e^{\frac{i\pi}{2\tau}(\alpha^2 - |\alpha|^2)} \frac{\theta(\alpha, \tau)}{\eta(\tau)}, \qquad \det(D^\dagger D) = |\det D|^2,$$

where

$$\theta(\alpha, \tau) = \sum_{n=-\infty}^{+\infty} e^{i\pi\tau n^2 + i 2\pi n \alpha} \qquad \eta(\tau) = e^{\frac{i\pi\tau}{12}} \prod_{n=1}^{\infty} (1 - e^{i 2\pi\tau n}).$$

Note the appearance of $|\alpha|^2$, the term violating holomorphic factorization of $\det(D^\dagger D)$. Our choice differs from that of ref. [2] by a pure phase prefactor. This is a consequence of imposing invariance under a 90 degree rotation (item 1).

We now imagine replacing the continuum torus by a toroidal square lattice consisting of an $L_1 \times L_2$ rectangle with $\frac{L_1}{L_2} = -i\tau$. On the links of the lattice we put a uniform lattice gauge field $U_\mu = e^{iaA_\mu}$ where the lattice spacing $a$ is $a = \frac{l_1}{L_1} = \frac{l_2}{L_2}$. The overlap prescription for the regularized $\det D$ is

$$O(\alpha, \tau, a) = \frac{{}_1\!<L-|L->_U}{|{}_1\!<L-|L->_U|} \frac{{}_U\!<L-|L+>_U}{{}_1\!<L-|L+>_1} \frac{{}_U\!<L+|L+>_1}{|{}_U\!<L+|L+>_1|},$$

where the states $|L\pm>_U$ are the ground states of many body Hamiltonians

$$H^\pm = \sum_{p,\alpha,\beta} a^\dagger(p,\alpha) h^\pm_{\alpha,\beta}(p;U) a(p,\beta) \qquad \{a^\dagger(p,\alpha), a(q,\beta)\} = \delta_{\alpha,\beta} \delta_{p,q}.$$

$\alpha, \beta = 1, 2$ and $p_\mu = \frac{2\pi}{L_\mu}(n_\mu + \frac{1}{2})$, $n_\mu = 0, 1, 2, \ldots, L_\mu - 1$. The single particle hermitian hamiltonians $h^\pm$ are given by

$$h^\pm(p; U) = \begin{pmatrix} \frac{1}{2}\hat{\Pi}^2 \mp m & i\tilde{\Pi} \\ -i\tilde{\Pi}^* & -\frac{1}{2}\hat{\Pi}^2 \pm m \end{pmatrix},$$

with

$$\tilde{\Pi}_\mu = \sin(p_\mu + aA_\mu), \quad \tilde{\Pi} = \tilde{\Pi}_1 + i\tilde{\Pi}_2 \quad \text{and} \quad \hat{\Pi}_\mu = 2\sin\frac{p_\mu + aA_\mu}{2}, \quad \hat{\Pi}^2 = \hat{\Pi}_1^2 + \hat{\Pi}_2^2.$$

The parameter $m$ can be chosen at will in the interval $(0, 1)$; we used the value $m = 0.9$. There is no holomorphy in $A$ in the regularized expression.

Our objective is to show that

$$\lim_{a \to 0} O(\alpha, \tau, a) = \frac{\det D(\alpha, \tau)}{\det D(0, \tau)} = e^{\frac{i\pi}{2\tau}(\alpha^2 - |\alpha|^2)} \frac{\theta(\alpha, \tau)}{\theta(0, \tau)}.$$

We do this by using the computer; the task is trivial numerically and a sample is shown in Figure 1.

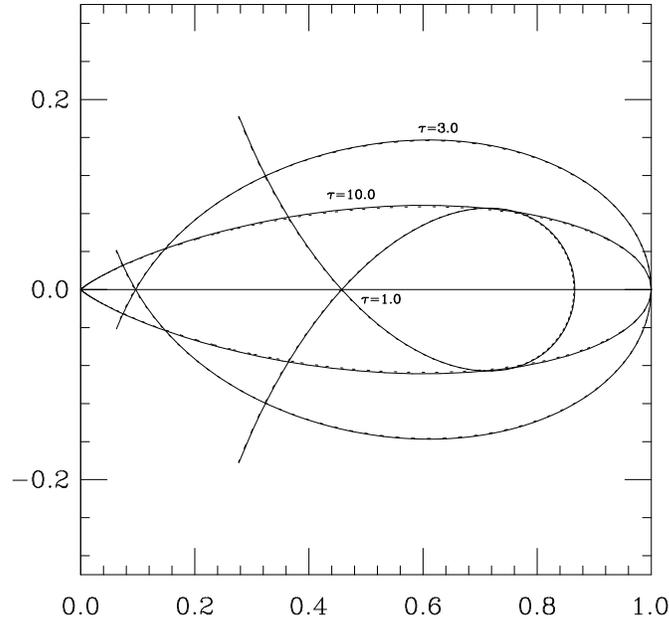

**Figure 1** The motion of the chiral determinant in the complex plane as a function of $-0.5 \leq \frac{l_2 A_2}{2\pi} \leq 0.5$ at fixed $\frac{l_1 A_1}{2\pi} = 0.37$ for three different values of $\tau$. The dashed curve represents the regularized quantity $O$, and the almost superimposed solid curve represents the corresponding continuum limit. $O$ was evaluated on a $30 \times 30$ ($\tau = 1$), $18 \times 54$ ($\tau = 3$) and $10 \times 100$ ($\tau = 10$) lattice. The lack of periodicity in $\frac{l_2 A_2}{2\pi}$ is reflected by the curves not closing.

It is easy to check that the lack of periodicity in the phase of $O$ (a consequence of the anomaly) cancels out from the product

$$\prod_{q_L} O(q_L \alpha, \tau, 0) \prod_{q_R} [O(q_R \alpha, \tau, 0)]^* =$$
$$\prod_{q_L} \frac{\theta(q_L \alpha, \tau)}{\theta(0, \tau)} \prod_{q_R} \left[\frac{\theta(q_R \alpha, \tau)}{\theta(0, \tau)}\right]^* e^{\frac{i\pi}{2\tau}[\sum q_L^2 (\alpha^2 - |\alpha|^2) - \sum q_R^2 (\alpha^{*2} - |\alpha|^2)]}$$

in the anomaly free case $\sum q_L^2 = \sum q_R^2 \equiv Q^2$. The exponential factor above becomes then $e^{\frac{i\pi}{2\tau} Q^2 (\alpha - \alpha^*)^2}$ and holomorphic factorization does not hold. The prefactor is real and positive in the anomaly free case and then only the $\theta$–functions contribute to the phase. Since all of the renormalization related choices affect only the prefactor, the phase is scheme independent.

Why is this test interesting ? The most characteristic feature of a four dimensional anomaly–free chiral gauge theory is reflected by parity violating interaction terms between gauge bosons in the fermion induced action. In the continuum limit a large amount of universal "strictly chiral" information is contained in the phase of the chiral determinant and a correct regularization should reproduce this universal content. In the two dimensional abelian case, where the overlap is relatively easy to compute, all nonzero modes of the gauge fields enter only through the bubble diagram. The reason for this is holomorphy combined with gauge invariance: A diagram with three or more legs is convergent and thus must be a function of only $A$ by holomorphy and unchanged under $A \to A + \partial_z \Lambda$ by gauge invariance. Hence, in the zero topological sector, all dependence on $A$ at powers higher than 2 disappears except through the zero mode $\int d^2 x A$ (more precisely, only a dependence on the gauge invariant part of the latter is allowed). Anomaly cancellation eliminates the imaginary part of the fermion induced action coming from the bubble, which escapes the above arguments needing regularization. Thus, in two dimensional abelian anomaly free models the single universal parity breaking terms in the chiral determinant are those coming from the fermion loop with all external legs at zero momentum. Our check has just shown that these terms are correctly reproduced by the overlap. In particular, Figure 2 shows an example of parity violating gauge invariant terms in the abelian anomaly free 345 model consisting of two left handed fermions of charge 3 and 4 and one right handed fermion of charge 5.

Having string applications in mind we suggest repeating the above for an orientable surface of genus two say, using random two dimensional lattices, generated, for example, from "large" Feynman diagrams of a matrix model.

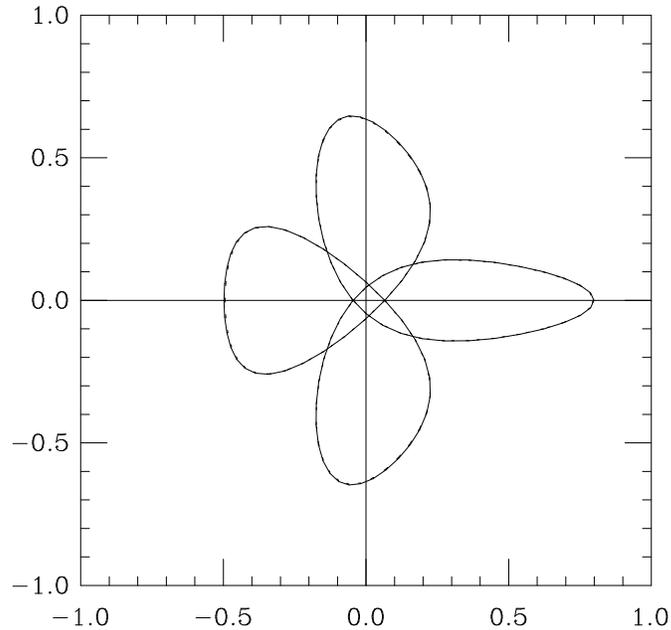

**Figure 2**  The motion of the chiral determinant of the 345 model in the complex plane as a function of $-0.5 \leq \frac{l_2 A_2}{2\pi} \leq 0.5$ at fixed $\frac{l_1 A_1}{2\pi} = 0.37$ at $\tau = 1$. The solid curve and the dashed curve have the same meaning as in Figure 1. The regularized expression was evaluated on a $30 \times 30$ lattice. Since gauge invariance is restored the curve closes now.

**Acknowledgments**: The research of R. N. was supported in part by the DOE under grant # DE-FG02-90ER40542 and that of H. N. was supported in part by the DOE under grant # DE-FG05-90ER40559.

**References**

[1] R. Narayanan, H. Neuberger, IASSNS-HEP-94/99, hep-th #9411108.
[2] L. Alvarez–Gaume, G. Moore, C. Vafa, Comm. Math. Phys. 106 (1986) 1.